\documentclass[prl,preprint,showpacs]{revtex4}
\usepackage{epsf}
\usepackage{psfig}
\usepackage{graphicx}
\begin{document}
\title{
Influence of Ni doping on the electronic structure of  Ni$_2$MnGa} 

\author{Aparna~Chakrabarti$^1$, C.~Biswas$^2$, S.~Banik$^2$, R.~S.~Dhaka$^2$, 
A.~K.~Shukla$^2$, and S.~R.~Barman$^{2*}$}

\affiliation{$^1$Solid State Laser Division, Centre for Advanced Technology, 
Indore, 452013, M.P., India} 
\affiliation{$^2$UGC-DAE Consortium for Scientific Research, University Campus, Khandwa Road, Indore, 452017, M.P., India}

\begin{abstract}
The modifications in the electronic structure of Ni$_{2+x}$Mn$_{1-x}$Ga by Ni doping 
have been studied using full potential linearized augmented plane wave method and ultra-violet photoemission spectroscopy. Ni 3$d$ related electron states appear due to formation of Ni clusters. We show the possibility of changing the minority-spin DOS with Ni doping, while the majority-spin DOS remains almost unchanged.
The total magnetic moment decreases with excess Ni. The total energy calculations corroborate the experimentally reported changes in the Curie temperature and the martensitic transition temperature with $x$.   
\end{abstract}
\maketitle

Ni$_2$MnGa is an unique material because it exhibits both ferromagnetism and shape memory effect and is an ideal candidate for magnetically controlled shape memory applications\cite{Murray00,Sozinov02}. Highest known magnetic field induced strain, giant magnetocaloric effect and large negative magnetoresistance have been reported in Ni-Mn-Ga\cite{Murray00,Sozinov02,Marcos03,Zhou04,Biswas05}. Physical properties of Ni$_2$MnGa are highly composition dependent. It is reported that replacing Mn by  Ni in Ni$_{2+x}$Mn$_{1-x}$Ga for $x$=~0 to 0.2 causes  the  Curie temperature ($T_C$) to decrease from 376 to 325~K and the martensitic transition temperature ($T_M$) to increase from 210 to 325~K\cite{Vasilev99,Zuo99}. 
These interesting properties make Ni$_{2+x}$Mn$_{1-x}$Ga a very important system for both fundamental physics and technological applications.  

The explanation of the above mentioned characteristics of Ni doped Ni$_2$MnGa is related to its electronic structure. So, in this letter, we investigate the electronic structure of Ni$_{2+x}$Mn$_{1-x}$Ga  using ultra-violet photoemission spectroscopy (UPS) and full potential linearized augmented plane wave (FPLAPW) calculations. Different band structure studies on Ni$_2$MnGa and related 
Heusler alloys in literature deal with the stoichiometric composition\cite{Kubler83,Fujii89,Lin92,Veliko99,Ayuela99,Bungaro03,Zayak03}; but very few study the effect of  Ni doping\cite{Enkovaara03,MacLaren02}. Recently, Enkovaara {\it et al.} showed that in Mn rich Ni$_2$MnGa, the doped Mn atoms are antiferromagnetically aligned\cite{Enkovaara03}. MacLaren calculated the density of states (DOS) for 20\% Ni rich Ni$_2$MnGa using layer KKR method and  correlated the structural properties with DOS\cite{MacLaren02}.

The ab-initio relativistic spin-polarized FPLAPW calculations were performed using WIEN97 code\cite{Wien97} with generalized gradient approximation\cite{Perdew96} for exchange correlation. An energy cut-off for the plane wave expansion of 16~Ry and $l_{max}$=~10 were used.  The muffin-tin radii were taken to be : Ni 1.19~\AA,~Mn 1.27~\AA,~Ga 1.19~\AA. The number of $k$ points in the irreducible  Brillouin zone (BZ) for self-consistent field cycles and DOS calculation varied between 147-168 for different structures. For the $x$=~0 tetragonal martensitic phase, the calculations were performed with the experimentally determined lattice constants: $a$=~$b$=~5.92~\AA,~ $c$=~5.56~\AA, space group $Fmmm$ ($Z$=~4) with atomic positions $8f$ (Ni), $4b$ (Mn) and $4a$ (Ga) without considering modulation\cite{Webster84,Wedel99}. FPLAPW calculation has also been performed with the real  structure with seven layer modulation\cite{Brown02} in $P_{nnm}$ space group ($Z$=~14) with $a$=~4.215, $b$=~29.302 and c=~5.557\AA~ with 80 $k$ points in the irreducible BZ, and is indicated in text as $x$=~0$M$. 
For Ni$_{2.25}$Mn$_{0.75}$Ga, our x-ray diffraction (XRD) studies do not show any modulation\cite{Banik05} and this is in agreement with literature\cite{Pons00}.
So, for Ni excess Ni$_{2.25}$Mn$_{0.75}$Ga, calculations were performed by replacing a Mn by Ni in the non-modulated $Fmmm$ structure and with the same lattice constants as $x$=~0 (henceforth designated as $x$=~0.25). Calculation was also performed in same structure but with the actual lattice constants of Ni$_{2.25}$Mn$_{0.75}$Ga ($a$=~$b$=~5.439\AA,~$c$=~6.563\AA, and $c/a$=~1.2)\cite{Banik05}. This is henceforth referred to as $x$=~0.25(1.2). The number 1.2 in bracket is the $c/a$ value.  All the DOS calculations shown here are done in the ferromagnetic ground state.\cite{Biswas05,Vasilev99,Zuo99,Albertini02}

Polycrystalline ingots of Ni$_{2+x}$Mn$_{1-x}$Ga were prepared by standard procedure\cite{Vasilev99}. Our XRD, differential scanning calorimetry (DSC), resistivity results agree with literature\cite{Vasilev99,Wedel99,Webster84,Pons00}. Energy dispersive analysis of x-rays shows that the samples are homogeneous. The
intended and actual compositions agree well, {\it e.g.} Ni$_{2.02}$Mn$_{0.97}$Ga$_{1.02}$ and Ni$_{2.21}$Mn$_{0.78}$Ga$_{1.01}$ for $x$=~0 and 0.2, respectively. He~{\small I} UPS (h$\nu$=~21.2~eV) was performed at a base pressure of $6\times10^{-11}$ mbar  using an   electron energy analyzer from Specs GmbH, Germany. The samples were mechanically  scraped to expose fresh surfaces devoid of  oxygen and carbon contamination. 

The atomic photoemission cross-sections of Ni 3$d$ and Mn 3$d$ at h$\nu$=~21.2~eV are 4.0 and 5.3~mega barn, respectively\cite{Yeh}. These values are an 
order of magnitude higher than that of Ni 4$s$, Mn 4$s$, and Ga 4$s, p$ states\cite{Yeh}. So,  the UPS valence band (VB) spectra are calculated by adding the partial DOS (PDOS) of  Ni and Mn 3$d$ , multiplied by their respective photoemission cross-sections\cite{Brown98}. This added DOS is multiplied with the Fermi function at the measurement temperature and  convoluted with a Voigt function. The FWHM of the Gaussian component ($\approx$100~meV) of the Voigt function represents the instrumental resolution. The energy dependent Lorentzian FWHM that represents the life-time broadening is 0.3$E$, where $E$ is the energy  {\it w.r.t.} $E_F$\cite{Fujimori84}. 

We compare the experimental UPS spectrum of Ni$_2$MnGa in the martensitic phase with the calculated VB  of  both non-modulated ($x$=~0) and modulated ($x$=~0$M$) structures in Fig.~1. 
\begin{figure}[htb]
\epsfxsize=150mm
\epsffile{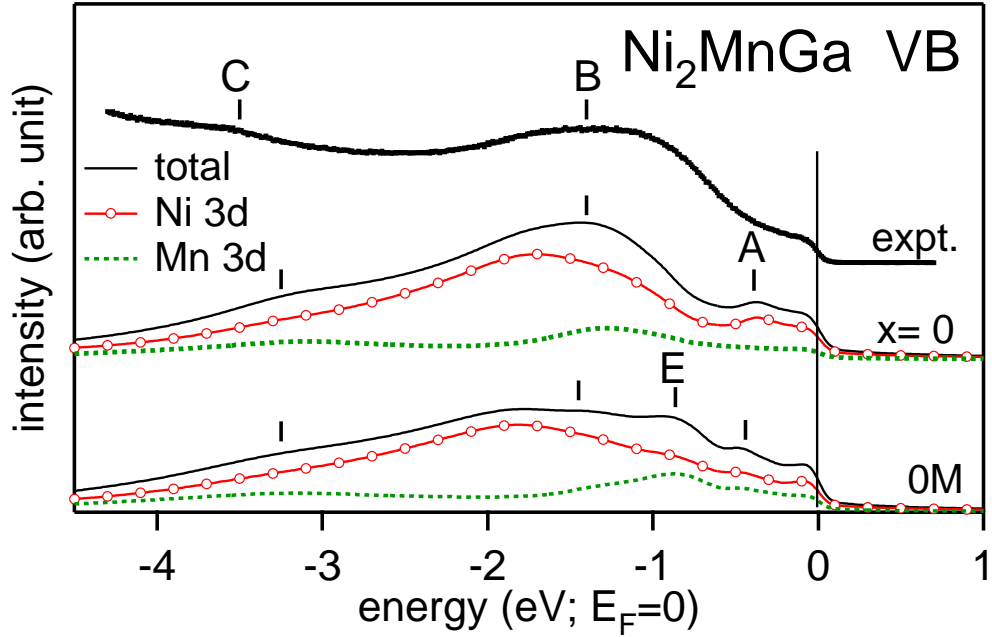}
\caption{
(Color online) Valence band (VB) spectra of Ni$_2$MnGa ($x$=~0) in the martensitic phase from
He~{\small I} ultra-violet photoemission (black dots) and FPLAPW calculation (black solid line). Ni 3$d$ (red solid line with open circles) and Mn 3$d$ (green dashed line) contributions to the calculated VB are shown. The spectra are staggered.}
\label{fig1}
\end{figure}
The UPS spectrum exhibits a broad main peak centered at -1.4~eV ($B$) and a weak feature at -3.5~eV ($C$). The cut-off at 0~eV is the Fermi level ($E_F$). The shape of $B$ (corresponding DOS feature is $B'$ in Fig.~2b) of  $x$=~0 is in good agreement with the UPS spectrum. The VB is largely dominated by Ni 3$d$ states with peak at -1.75~eV. Mn 3$d$ states exhibit two features at -1.3 and~~ -3.1~eV. Feature $C$  arises due to Ni 3$d$- Mn 3$d$ bonding and is related to feature $C'$ in the DOS (Fig.~2b).
Features $A$, $B$, and $C$ appear at similar energies in $x$=~0 and 0$M$. This is expected because, although 0$M$ has large orthorhombic unit cell, modulation involves a small (0 to 5\%) periodic shuffle of the (110) atomic planes of the  non-modulated structure along the [1$\overline{1}$0] direction\cite{Martynov92}. However, in 0$M$ VB, a  feature $E$ appears prominently, which has hardly any signature in the experimental spectrum. This feature is related to Mn 3$d$- Ni 3$d$ hybrid states that are more intense and appear at -0.9~eV, in contrast to $x$=~0 VB. It is evident from Fig.~1 that $x$=~0 VB is in better agreement with experiment than 0$M$ VB. The possible reasons could be : (i) mean free path of photoelectrons in UPS is 10-15~\AA,~ hence UPS probes the 0$M$ structure ($b$$\approx$~29~\AA)~ only partially; (ii) due to possible surface relaxation or reconstruction effects, the modulated structure is modified or absent at the surface.  The Ni 3$d$ related feature $A$ at  -0.4~eV (feature $A'$ in Fig.~2b) is absent in the experiment. Similar discrepancy between UPS and calculated VB has been reported for other Mn based Heusler 
\begin{figure}[htb]
\epsfxsize=150mm
\epsffile{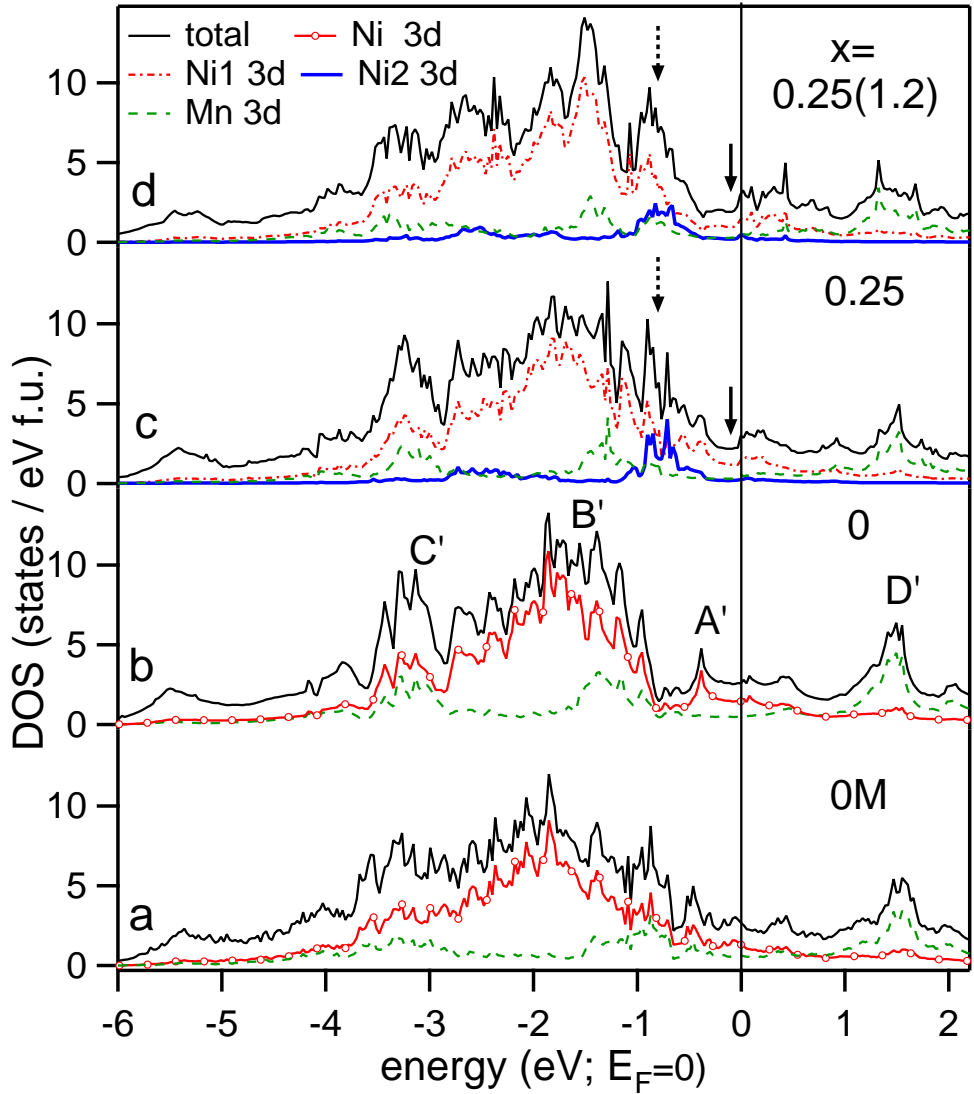}
\caption{(Color online) Total DOS (black solid line) and Ni 3$d$ (red solid  line with open circles for (a) and (b); red dot-dashed line and blue thick solid line for (c) and (d), see text) and Mn 3$d$ (green dashed line) partial DOS of ferromagnetic Ni$_{2+x}$Mn$_{1-x}$Ga in the martensitic phase.}
\label{fig2}
\end{figure}
alloys\cite{Brown98} and the possible reasons are discussed later.  

To investigate the effect of Ni doping only, we compare the $x$=~0 and 0.25 DOS in Fig.~2(b, c), both  with lattice constants of $x$=~0.  In  both cases, the DOS is dominated by Ni and Mn 3$d$ bonding states ($B'$ and $C'$), as in other related Heusler alloys\cite{Kubler83,Lin92}. The occupied Mn 3$d$ PDOS is split into clearly separated $e_g$ and $t_{2g}$ states,  appearing at -1.3 and -3~eV, respectively. Feature $D'$ in the anti-bonding region above $E_F$ is largely dominated by Mn 3$d$ PDOS. Interesting difference between $x$=~0 and 0.25 is observed around -0.8~eV, where  new electron states appear in the latter (dashed arrow) that fill up the valley between $A'$ and $B'$ in $x$~=0. The PDOS of Ni2, {\it i.e.} the doped  Ni atom in Mn position, has its maximum at -0.8~eV. 3$d$ PDOS of Ni1 (Ni atoms in Ni position) also appear at this energy with similar intensity. This shows that these states arise from the bonding between doped and existing Ni atoms. In fact, since the  nearest neighbor (n.n.) of a Mn atom are 8 Ni atoms, when Mn is replaced by Ni, a  9 atom body centered tetragonal  Ni cluster is formed with n.n. distance 2.53\AA. Hence, Ni-Ni  bonding occurs at the expense of Ni-Mn bonding. To test this explanation, we have calculated the DOS of Ni$_2$Mn$_{0.75}$Ga$_{1.25}$ where, instead of Ni, Ga replaces Mn. As expected, the DOS does not show the Ni related extra states because in this case Ni clustering does not occur\cite{Chakrabarti04}. 

The change in electronic structure in the Ni doped case between the real structure [$x$=~0.25(1.2), $c/a$=~1.2] and non-equilibrium $x$=~0.25 structure ($c/a$=~0.94) is shown in Fig.~2(c,d). The total energy ($E_{tot}$) calculated using FPLAPW confirm that 0.25(1.2) is more stable than 0.25 by 12.2 meV/atom.
In 0.25(1.2), the Ni 3$d$ related new states appear as a clear peak at -0.8~eV. The Ni1 3$d$ states are more intense than the Ni2 3$d$ states.  The Mn 3$d$ related feature $D'$ is broadened.  Comparison of Fig.~2(a,d) shows the difference between the DOS of undoped (0$M$) and Ni doped Ni$_2$MnGa [0.25(1.2)], both with actual experimentally determined structures. In 0$M$, Ni~3$d$-Mn~3$d$ hybrid states are observed around 0.9 eV that gives rise to feature $E$ in Fig.~1. In contrast, in 0.25(1.2), Ni1~3$d$- Ni2~3$d$ hybrid states dominate this region. 

 A  dip is observed in the $x$=~0.25 DOS in the near $E_F$ region between $A'$ and $E_F$ (black arrow, Fig.~2). To understand its origin, the spin projected DOS in the near $E_F$ region is shown in Fig.~3a. We find that the dip is due to a decrease in 
\begin{figure}[htb]
\epsfxsize=150mm
\epsffile{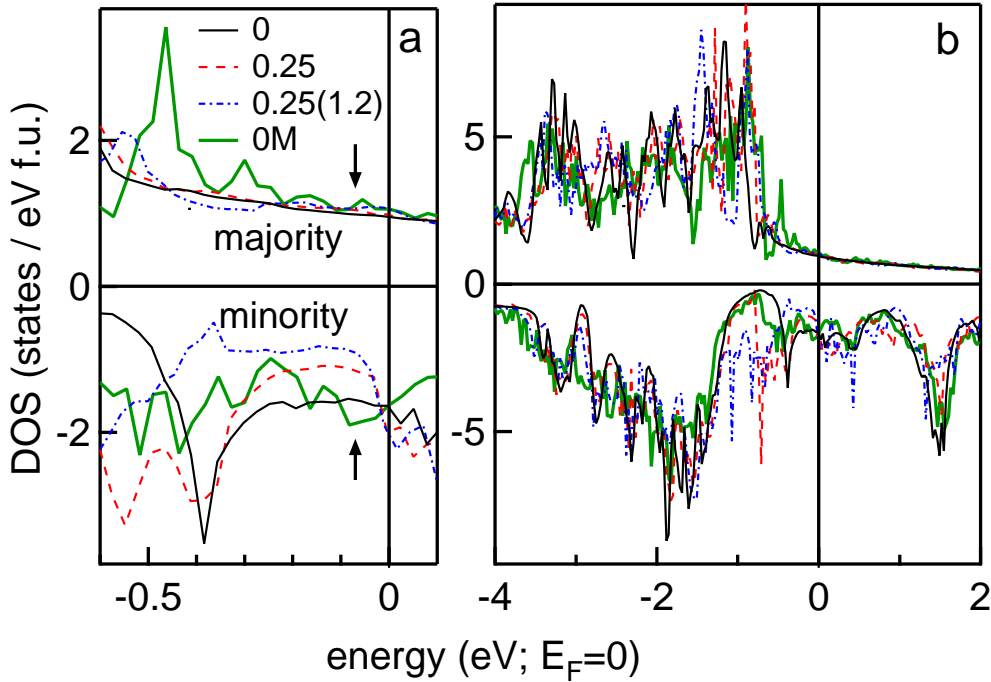}
\caption{(Color online) Spin-projected DOS of Ni$_{2+x}$Mn$_{1-x}$Ga (a) in near $E_F$ (b) extended region for $x$=0 (black solid line), 0.25 (red dashed line), 0.25(1.2) (blue dot-dashed line) and 0M (green thick solid line).}
\label{fig3}
\end{figure}
the minority-spin DOS near $E_F$ by 31\% in $x$=~0.25 {\it w.r.t.} $x$=~0 (from 1.6 to 1.1 states/eV f.u. at -0.15 eV). If compared to 0$M$, the decrease even larger (41\%). Because of Ni doping, there is a redistribution of the partially filled minority-spin DOS.  Both Ni and Mn 3$d$ PDOS decrease in the near $E_F$ region resulting in the dip, while new Ni1- Ni2 hybrid states appear around -0.8~eV. 
The minority-spin DOS is further reduced by 43\% in 0.25(1.2) compared to $x$=~0, and {\it w.r.t.} 0$M$ the decrease is 55\% (Fig.~3a). In 0.25(1.2), feature $A'$ is absent and the dip is more pronounced and broadened (-0.55~eV to $E_F$). In contrast, the majority-spin DOS near $E_F$ remains essentially unchanged in all cases  (arrow, Fig.~3a), because they are almost fully filled\cite{MacLaren02}. Thus, Ni$_{2+x}$Mn$_{1-x}$Ga presents an exciting possiblity of tuning  the minority-spin DOS near $E_F$ with Ni doping, which might have interesting implications in spin polarized transport.
Comparing Figs.~2 and 3b, we find that $D'$ originates predominantly from Mn 3$d$ minority-spin states, while Mn 3$d$ majority-spin states are almost fully occupied. 
This is the reason why Mn has large magnetic moment. Features $B'$ and $A'$ are dominated by minority-spin states in all cases, except for 0$M$ where $A'$ is broader and has similar 
contributions from both spin states.  Feature $C'$ is dominated by majority-spin states. 

UPS VB spectra in Fig.~4 show that the main peak of Ni$_{2.1}$Mn$_{0.9}$Ga 
\begin{figure}[htb]
\epsfxsize=150mm
\epsffile{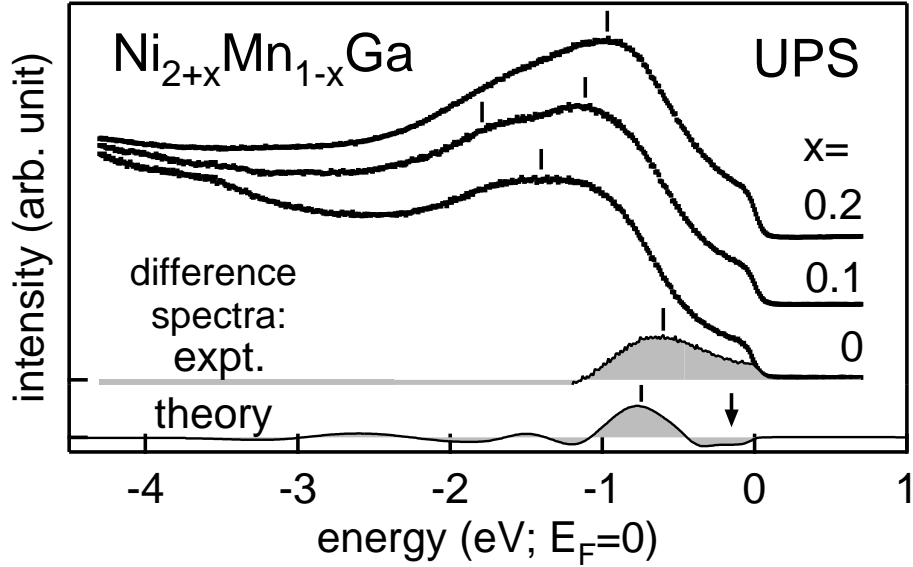}
\caption{
UPS spectra of Ni$_{2+x}$Mn$_{1-x}$Ga in the martensitic phase. The spectra have been normalized to the same height and staggered along the vertical axis.  The experimental (between $x$=~0.2 and 0) and calculated (between  $x$=~0.25(1.2) and  $x$=~0) difference spectra are shown.}
\label{fig4}
\end{figure}
($x$=~0.1) is centered at -1.1~eV with a  shoulder at -1.65~eV, in contrast to Ni$_2$MnGa. For Ni$_{2.2}$Mn$_{0.8}$Ga ($x$=~0.2), the -1.1~eV peak becomes relatively more intense and shifts to -0.9~eV.  The experimental difference spectrum between $x$=~0.2 and 0 shows extra states in the former around -0.65~eV. The difference spectrum from theory between $x$=~0.25(1.2) and 0 exhibits a peak at -0.75~eV in agreement with experiment and these extra states are the new Ni1-~Ni2 3$d$ bonding states, as discussed earlier (Fig.~2). The agreement is not good if the  0$M$ VB is used to calculate the theoretical difference spectrum  (not shown in Fig.~4) and this supports our contention that the surface electronic structure is hardly influenced by modulation. The dip in the DOS between -0.4 eV and $E_F$ is observed in the theoretical difference spectrum (arrow, Fig.~4), but is not clearly observed from experiment. Possible reasons for this disagreement could be surface relaxation or presence of anti-site defects\cite{Larson00}, {\it etc}. To find the effect of surface relaxation, we have calculated the $x$=~0.25 DOS with 10\% lattice expansion.
~We find that the dip region shifts to $E_F$\cite{Chakrabarti04}, which would imply absence of the dip below $E_F$ in UPS. Also, absence of feature $A$ (Fig.~1) in UPS is explained by the shift in the DOS. We have calculated the DOS of $x$=~0.25 with a simple anti-site defect (Ni and Mn positions interchanged). 
~The DOS does not show the dip and feature $A'$ is absent.  
 Thus, surface relaxation and/or anti-site defects are likely to be responsible for the present and similar earlier\cite{Brown98} discrepancies between experiment and theory. 

Now we turn to the discussion of the bulk magnetic moments calculated using FPLAPW for the real structures. The total magnetic moment for Ni$_2$MnGa (0$M$) is 3.81~$\mu_B$, and the local moments per site for Mn, Ni, and Ga are   3.06, 0.21, and -0.03~$\mu_B$, respectively.
~These magnetic moments are in better agreement with experiment\cite{Brown99} compared to $x$=~0 (total:~4.13, Mn:~3.44, and Ni:~0.36~$\mu$B). Thus, magnetic moment, which is a bulk property, is better described by the 0$M$ structure.  For $x$=~0.25(1.2), the total magnetic moment is 3.31~$\mu_B$,
~and the local moments per site for Ni1, Ni2, Mn and Ga are 
0.37, 0.23, 3.41, and -0.03~$\mu_B$,
~respectively. The doped Ni2 is in ferromagnetic configuration, and its moment is smaller than Ni1.  Although the Mn moment increases with Ni doping, the total moment decreases because the Ni2 moment is less than that of Mn it replaces. Magnetization measurements also show a decrease in total saturation moment with excess Ni\cite{Albertini02}. 
 
For Ni$_2$MnGa, the ferromagnetic (FM) transition occurs in the austenitic phase since $T_C$$>$$T_M$. So, to find the stability of the FM state, difference in $E_{tot}$ between the paramagnetic (PM) and the FM state ($\Delta$$E_{tot}$) has been calculated in the austenitic phase. We find  $\Delta$$E_{tot}$ to be 322
~meV/atom for Ni$_2$MnGa. In contrast, our recent DSC studies on  Ni$_{2.25}$Mn$_{0.75}$Ga show that $T_M$$>$$T_C$ and the magnetic transition occurs in the martensitic phase,\cite{Banik05} and $\Delta$$E_{tot}$ turns out to be 219 meV/atom.
~Since Ni$_2$MnGa satisfies Stoner condition of ferromagnetism, it is possible to approximately relate $T_C$ to $\Delta$$E_{tot}$ by: $k_BT_C$$\sim$$\Delta$$E_{tot}$\cite{Fujii89,Veliko99}. 
Although this  expression  gives a much higher value of $T_C$\cite{Veliko99},
it could be used to find the relative variation. Thus, w.r.t. $T_C$ (=~376~K) of $x$=~0;  for 0.25(1.2) $T_C$ should be 256~K.
~So, from theory we find that $T_C$ decreases with Ni doping, which explains the experimental data\cite{Vasilev99,Zuo99,Albertini02}.

In order to relate $T_M$ to $x$,  we note that $E_{tot}$ for the martensitic phase should be lower than the austenitic  phase, since the former is the lower temperature phase. Higher total energy difference between the two phases ($\delta$$E_{tot}$) would imply greater stability of the latter and hence enhanced $T_M$. For NiTi and PdTi this is indeed found to be so\cite{Bihlmayer96}. For Ni$_2$MnGa in FM state, $\delta$$E_{tot}$ is 
3~meV/atom.
~For $x$=~0.25(1.2) in PM state, $\delta$$E_{tot}$ is 39~meV/atom.
Thus, the martensitic  phase is more stable compared to the austenitic phase in 0.25(1.2) {\it i.e.} the Ni doped case. This is consistent with the experimentally observed higher $T_M$ with Ni doping\cite{Vasilev99,Zuo99,Banik05}. 

In conclusion, based on FPLAPW calculations and photoemission spectroscopy, we show that with Ni doping new Ni related electron states appear due to formation of Ni clusters in Ni$_{2+x}$Mn$_{1-x}$Ga.  The reported trends in the variation of  $T_C$, $T_M$ and magnetic moments with $x$ are explained. For Ni$_2$MnGa, the effect of modulation in the structure is not evident at the surface, while bulk property like magnetic moment is better described by the modulated 0$M$ structure. Although not clearly observed in the spin integrated UPS spectra possibly due to surface relaxation or anti-site defects, we find that with Ni doping a dip appears below $E_F$ in the minority-spin DOS calculated using FPLAPW. In contrast, the  majority-spin DOS remains unchanged. This indicates a probable future application of Ni$_{2+x}$Mn$_{1-x}$Ga in spin polarized transport with tunable efficiency through controlled doping.

We thank Mr. T. P. S. Nathan,  Dr. P. Chaddah, and Professor A. Gupta for constant  encouragement. Dr. N. P. Lalla, Dr. A. M. Awasthi and Dr. R. Rawat are thanked for providing sample characterization facilities; and Mr. S. C. Das is thanked for technical help. SB and AKS acknowledge the financial support from D.S.T.
 
\newpage
\noindent *email: srbarman@mailcity.com, barman@udc.ernet.in

\newpage
\noindent {\bf Figure Captions}

\noindent Figure 1:~ (Color online) Valence band (VB) spectra of Ni$_2$MnGa ($x$=~0) in the martensitic phase from
He~{\small I} ultra-violet photoemission (black dots) and FPLAPW calculation (black solid line). Ni 3$d$ (red solid line with open circles) and Mn 3$d$ (green dashed line) contributions to the calculated VB are shown. The spectra are staggered.\\

\noindent Figure 2:~(Color online) Total DOS (black solid line) and Ni 3$d$ (red solid line with open circles for (a) and (b); red dot-dashed line and blue thick solid line for (c) and (d), see text) and Mn 3$d$ (green dashed line) partial DOS of ferromagnetic Ni$_{2+x}$Mn$_{1-x}$Ga in the martensitic phase.\\

\noindent Figure 3:~(Color online) Spin-projected DOS of Ni$_{2+x}$Mn$_{1-x}$Ga (a) in near $E_F$ (b) extended region for $x$=0 (black solid line), 0.25 (red dashed line), 0.25(1.2) (blue dot-dashed line) and 0M (green thick solid line).\\

\noindent Figure 4:~UPS spectra of Ni$_{2+x}$Mn$_{1-x}$Ga in the martensitic phase. The spectra have been normalized to the same height and staggered along the vertical axis.  The experimental (between $x$=~0.2 and 0) and calculated (between  $x$=~0.25(1.2) and $x$=~0) difference spectra are shown.\\

\end{document}